\documentclass[twoside]{article}
\usepackage{fleqn,espcrc2}

\newcommand{\AmS}{{\protect\the\textfont2
  A\kern-.1667em\lower.5ex\hbox{M}\kern-.125emS}}

\title{\bf Free conical dynamics: charge-monopole as a
particle with spin,
anyon and nonlinear fermion-monopole supersymmetry}

\author{Mikhail S. Plyushchay\address[DF]{
Departamento de F\'{\i}sica,
Universidad de Santiago de Chile,
Casilla 307, Santiago 2, Chile}\address[P]{
     Institute for High Energy Physics, Protvino,
Russia}
        \thanks{E-mail: mplyushc@lauca.usach.cl}}

\makeatletter
\def\bsymb#1{
  \begingroup
  \let\@nomath\@gobble \boldmath
  \mathchoice
    {\hbox{$\m@th\displaystyle#1$}}
    {\hbox{$\m@th\textstyle#1$}}
    {\hbox{$\m@th\scriptstyle#1$}}
    {\hbox{$\m@th\scriptscriptstyle#1$}}
  \endgroup}
\makeatother

\def\cc#1{\kern .7em\hfill #1 \hfill\kern .7em}
\def\ZZ{\hbox{\it Z\hskip -4.pt Z}}

\def\RR{\hbox{\it I\hskip -2.pt R }}

\newcommand{\nc}{\newcommand}
\nc{\beq}{\begin{equation}}
\nc{\eeq}{\end{equation}}
\nc{\beqa}{\begin{eqnarray}}
\nc{\eeqa}{\end{eqnarray}}
\nc{\noi}{\noindent}

\begin{document}

\begin{abstract}
We discuss the origin of the  effectively free
dynamics of the charge  in the magnetic monopole field
to apply it for finding  the alternative treatment
of the charge-monopole as a particle with spin,
for tracing out  the  relation of the charge-monopole to
the free relativistic  anyon and  for clarifying   the nature of
the
non-standard nonlinear  supersymmetry of the
fermion-monopole system.
\vspace{1pc}
\end{abstract}

\maketitle

\section{Introduction}
In addition to the Galilei symmetry,
the free nonrelativistic particle possesses the
dynamical $SL(2,R)$ symmetry \cite{sl2r}.
The latter can be
understood as a relic of reparametrization invariance
which survives  the formal
Lagrangian gauge fixing procedure
consisting in introduction of the condition
$v=1$  in  the associated system
$L=\frac{\dot{\bf r}{}^2}{2v}+\frac{v}{2}$
with   einbien $v$ \cite{mp1}.
Amazingly, this dynamical
symmetry relates the mechanisms of supernova
explosion and plasma implosion \cite{astro},
and the theoretical explanation of such an intriguing
similarity finds its roots in the  symmetry properties of
the fluid dynamic equations \cite{fluid}.
Since the charge-monopole (CM) system is characterized
by the same dynamical $SL(2,R)$ symmetry \cite{jac},
this indicates on the  free nature
of the charge's motion  in such a system.
Following recent works  \cite{mp1,mp2}, here  we
discuss the origin of the  effectively free dynamics
of the charge  in the magnetic monopole field
to apply it for finding  the alternative treatment
of the CM  as a particle with spin,
for tracing out  the  relation of the CM to
the free relativistic  anyon, and  for clarifying   the nature of
the
non-standard nonlinear  supersymmetry of the
fermion-monopole system.

\section{Free  dynamics of the charge-monopole}
A non-relativistic scalar particle of unit mass and electric
charge $e$ in the field of magnetic monopole of charge $g$
is described by the Lagrangian
\beq
\label{lchmo}
L=\frac{1}{2}{\dot{\bf r}}{}^2+e{\bf A}{\dot{\bf r}},
\eeq
with a U(1)  gauge potential ${\bf A}({\bf r})$
given by the relations
$
\partial_iA_j-\partial_jA_i= F_{ij}=\epsilon_{ijk}B_k,$
$
B_i=g\frac{r_i}{r^3},
$
$
r=\sqrt{{\bf r}^2}
$.
Equations of motion following from Lagrangian
(\ref{lchmo})
result in the Lorentz force law,
$\ddot{\bf r}={\bf  f}$,
$
{\bf f}=-\nu r^{-3}
{\bf r}\times \dot{\bf r}
$,
$\nu=eg$,
which implies that instead of the orbital angular momentum
vector
$
{\bf L}={\bf r}\times {\dot{\bf r}},
$
the vector
\beq
\label{jint}
{\bf J}={\bf L}-\nu{\bf n},
\quad {\bf n}={\bf r}\cdot {r^{-1}},
\eeq
is the conserved  angular momentum of the system.
Due to the relation  ${\bf Jn}=-\nu$,
the trajectory of the particle lies on the cone
with the  axis  oriented along the
vector ${\bf J}$ and the  half-angle  equal to
$
\cos\gamma=-\nu J^{-1}.
$
Since the force ${\bf f}$ is orthogonal to
${\bf r}$ and to the velocity $\dot{\bf r}$,
it is perpendicular to the cone and,
therefore, the particle performs a
{\it free motion on the cone}.
To  observe this also at the level of Lagrangian,
we introduce  (local)
spherical coordinates,
${\bf n}={\bf n}(\vartheta,\varphi)$,  with axis
$\vartheta=0$ directed along
${\bf J}$.
Then the reduction of  (\ref{lchmo})
to  the level of the constant vector ${\bf J}$
results after transformation
$
t\rightarrow  t'=\alpha t,
$
$
\alpha=
\sqrt{1-\nu^2 J^{-2}},
$
in the Lagrangian
\beq
\label{conus}
L=L_\alpha +\frac{d}{dt}\left(C
{\varphi}\right),\, \,
L_\alpha=\frac{1}{2}(\alpha^{-2}\dot{r}{}^2+r^2
\dot{\varphi}{}^2),
\eeq
where
$C=e(2\pi)^{-1}\Phi$,
$
\Phi=-2\pi \alpha \nu^2 J^{-1}e^{-1}.
$
The second term being  a total derivative
is not important {\it classically},
whereas the Lagrangian  $L_\alpha$
describes  a free motion of the
particle on the cone given by the relations
$
x=r\cos\varphi,
$
$
y=r\sin\varphi,
$
$
z=r\sqrt{\alpha^{-2}-1},
$
$r>0,
$
$
0\leq \varphi< 2\pi,
$
with $0<\alpha<1$.
The total derivative   in (\ref{conus})
is the reduced form of the
CM interaction $e{\bf A}\dot{\bf r}$
having  the  nature of $(0+1)$-dimensional Chern-Simons
(CS) term which
gives rise to the Dirac quantization condition for
the parameter $\nu$.
Being, up to the constant,
the derivative of the topologically nontrivial
angular variable, it  corresponds
to the 2D term describing the interaction
of the charge $e$ with a singular point vortex
carrying the magnetic flux
$\Phi$ \cite{jackvort}.

The conical metric
$ds^2=\alpha^{-2}(dr)^2+r^2(d\varphi)^2$
corresponds to the metric produced by the point mass
\cite{genqm}, and one
can say that the classical motion of the charge in
the field of magnetic monopole
(reduced to the level ${\bf J}=const$)
is equivalent to the classical motion
of a particle in a planar gravitational
field of a point massive source
carrying simultaneously the magnetic flux
$
\Phi.
$

With the appropriate
choice of the origin of the system of coordinates,
the motion of a 3D free particle ($e=0$)
can be characterized by
the unit vector ${\bf n}$ and by the
conserved orbital angular momentum ${\bf L}$
supplemented with the
canonically conjugate scalars $r$
and
$p_r=r^{-1} {\bf pr}$, where ${\bf p}$ is a canonical
momentum.
Since ${\bf Ln}=0$, for a given ${\bf L}$
the particle's trajectory is in  the plane
orthogonal to the orbital angular momentum.
So, one concludes that classically the topological nature
of $(0+1)$-dimensional charge-monopole CS term
is manifested in changing the
global structure of the dynamics without distorting
its local free (geodesic) character:
the ``plane dynamical geometry" of the free particle ($e=0$)
is changed for the free ``cone dynamical geometry" of the
charged particle.

\section{CM as a reduced E(3) system}
Like a 3D free particle,
the CM system may be treated
as a reduced E(3) system.
To get such an interpretation,
we  pass over from the  Hamiltonian variables ${\bf r}$
and ${\bf P}={\bf p}-e{\bf A}$
to the set of variables ${\bf n}$, ${\bf J}$,
$r$ and $P_r={\bf Pr}\cdot r^{-1}$.
They have the following Poisson brackets:
$
\{r,P_r\}=1,
$
$
\{r,{\bf n}\}=\{r,{\bf J}\}=\{P_r,{\bf n}\}=
\{P_r,{\bf J}\}=0,
$
and
$
\{J_i,J_j\}=\epsilon_{ijk}J_k,
$
$
\{J_i,n_j\}=\epsilon_{ijk}n_k,
$
$
\{n_i,n_j\}=0.
$
Poisson brackets for variables $J_i$ and $n_i$
correspond to the
algebra of generators of the Euclidean group E(3) with
$J_i$ being a set of generators of rotations
and $n_i$ identified as generators
of translations. The quantities ${\bf n}^2$ and
${\bf Jn}$ lying in the center
of $e(3)$ algebra, $\{{\bf n}^2,n_i\}=\{{\bf n}^2,J_i\}=
\{{\bf nJ},n_i\}=\{{\bf nJ},J_i\}=0$,
are fixed in the present case by the relations
$
{\bf n}^2=1,
$
$
{\bf nJ}=-\nu.
$
In terms of the introduced variables,
the Hamiltonian of the system takes the form
\beq
\label{hame3}
H=\frac{1}{2}P_r^2+\frac{({\bf J}\times
{\bf n})^2}{2r^2}.
\eeq
Therefore, the CM system can be treated
as the E(3) system reduced by the conditions
$
{\bf n}^2=1
$
and
$
{\bf nJ}=-\nu
$
fixing the Casimir elements,
and supplemented by the independent canonically
conjugate variables $r$ and $P_r$.
It is the relation
$
{\bf nJ}=-\nu
$
that encodes the topological difference between
the CM and the 3D free particle cases:
for $\nu\neq0$, the space given by the spin
vector ${\bf J}$ is homeomorphic to
$\RR^3-\{0\}$, ($J>|\nu|$),
whereas for $\nu=0$ the corresponding space
$\RR^3$ is topologically trivial.

\section{Spin nature of CS term}
The integrand in action corresponding to
the charge-monopole interaction term
$\theta=e\dot{\bf r}{\bf A}({\bf r})dt$
can be treated as a differential one-form
$
\theta=e{\bf A}({\bf r})d{\bf r}
$
defined by the relation
\beq
\label{ch-s}
d\theta=\frac{\nu}{2r^3} \epsilon_{ijk}r_idr_j\wedge
dr_k.
\eeq
Since the right-hand side of Eq. (\ref{ch-s}) is
the gauge-invariant
curvature two-form,
$
d\theta=e{\cal F},
$
$
{\cal F}=\frac{1}{2}F_{ij}dr_i\wedge dr_j
$,
the gauge-non-invariant one-form $\theta$
has a sense of $(0+1)$-dimensional CS term.
The two-form (\ref{ch-s}) can be represented equivalently
as
$
d\theta=\frac{\nu}{2}\epsilon_{ijk}n_idn_j\wedge dn_k.
$
With the (local) parametrization by spherical coordinates,
${\bf n}={\bf n}(\vartheta,\varphi)$,
one finds that up to the constant factor
it  gives  the differential area of the two-sphere
and  via the Stokes theorem leads to  the
quantization of the CM coupling constant:
$2\nu=k$, $k\in \ZZ$ \cite{tor1}.

Defining the dependent variables
$
s_i=-\nu n_i,
$
the two-form $d\theta$ can be represented
equivalently as
\beq
\label{omega}
\omega_{s}=
d\theta=-\frac{1}{2{\bf s}^2}
\epsilon_{ijk}
s_ids_j \wedge ds_k,
\eeq
with
$
{\bf s}^2=\nu^2.
$
The two-form (\ref{omega}) is closed and nondegenerate,
and, so,  can be interpreted as a symplectic form
corresponding to the symplectic potential $\theta$.
If we drop out the kinetic term in
the CM action (that corresponds to
taking the charge's zero
mass limit, $m\rightarrow 0$, \cite{djt}),
one gets the Poisson brackets
$
\{s_i,s_k\}=\epsilon_{ijk}s_k,
$
which together with the relation
$
s_is_i=\nu^2
$
define the classical spin system
with fixed spin modulus.
Geometric quantization applied to such
a system (for the details see ref. \cite{tor1})
leads to the same Dirac quantization of the parameter
$\nu$,
$|\nu|=j$, $j=1/2,1,3/2,\ldots$,
and results in $(2j+1)$-dimensional
representation of $su(2)$
with classical relation
${\bf s}^2=\nu^2$ changed for the quantum relation
${\bf s}^2=j(j+1)$.
Introducing the  complex variable $z$
related
to the spherical angles via the stereographic projection
$
z=\tan \frac{\vartheta}{2}e^{i\varphi}
$
from the north pole, or via
$
z=\cot \frac{\theta}{2}e^{-i\varphi}
$
for the projection from the south pole,
the symplectic two-form is represented
in both cases as
\beq
\label{wzz}
\omega_{s}=2i\nu\frac{d\bar{z}\wedge d{z}}{(1+\bar
{z}z)^2}.
\eeq
Geometrically, the obtained spin system is
a  K\"ahler manifold with K\"ahler
potential
${\cal K}=2i\nu \ln(1+\bar{z}z)$:
$\omega_{s}=\frac{\partial^2}{\partial z\partial\bar{z}}
{\cal K}\cdot d\bar{z}\wedge d{z}$, $\bar{z}=z^*$.
Locally, in  spherical coordinates
the spin Lagrangian   is given by
$
L_{s}=\nu\cos\vartheta\dot{\varphi},
$
and in terms of global complex variable
it  takes the form
\beq
\label{lspin}
L_{s}=i\nu \frac{\bar{z}\dot{z}-\dot{\bar{z}}z}
{1+\bar{z}z}.
\eeq
The appearance of the two stereographic projections
for the spin system  reflects
the necessity to work in two charts
under treating the  CM system to escape the
problems with Dirac string singularities.
In terms of globally defined independent variables $z$,
$\bar{z}$
no gauge invariance left in the spin system
given by Lagrangian (\ref{lspin})
but it is hidden in a fibre bundle structure
reflected, in particular, in the presence of two charts.
Note also that since (\ref{lspin}) is a first order Lagrangian
and the corresponding Hamiltonian is equal to zero,
the CS action $S=\int \theta$
describes  the  {\it free} spin
system.

\section{CM as a particle with spin}

The spin nature of the charge-monopole CS
term and free character of the charge's dynamics
allow us to get the alternative description
for the CM system as a
free particle of fixed spin with translational
and spin degrees of freedom interacting via the helicity
constraint and the spin-orbit coupling term.
The corresponding equivalent form of  the
Lagrangian is
\beq
\label{lfspin}
L=\frac{1}{2}\dot{\bf r}{}^2-
\frac{1}{r^2}({\bf r}\times \dot{\bf r})\cdot
{\bf s}
-\lambda \cdot ({\bf rs}+\nu r)
+L_{s},
\eeq
where ${\bf s}=-\nu{\bf n}(z,\bar{z})$,
$L_{s}$ is given by Eq.  (\ref{lspin}),
$\lambda$ is a Lagrange multiplier,
and the coordinate vector   ${\bf r}$
and translation invariant variables
$z$, $\bar{z}$ should be treated  as
independent variables of the configuration space.
To see that the Lagrangian (\ref{lfspin})
describes the system equivalent to the initial
CM system, first we note that the variation in
$\lambda$ results in appearance of the helicity constraint
$
\chi \equiv {\bf sr} + \nu r\approx 0,
$
whereas the term $L_{s}$ generates
the necessary Poisson brackets
$\{s_i,s_j\}=\epsilon_{ijk}s_k$
for the spin variables $s_i=s_i(z,\bar{z})$.
The Hamiltonian corresponding to the Lagrangian
(\ref{lfspin}) is
\beq
\label{ht}
H=\frac{1}{2}{\bf \Pi}^2 +\lambda \cdot ({\bf sr}+ \nu r),
\eeq
where
$
\Pi_i\equiv p_i-\frac{1}{r^2}\epsilon_{ijk}r_j s_k,
$
and it is the presence of the second ${\bf Ls}$-coupling
term in Lagrangian (\ref{lfspin}) in addition to the
first kinetic term that guarantees the conservation  of the
helicty  constraint.
The system
(\ref{lfspin}) is described by  $6+2$
phase space variables
and by one first class constraint.
Therefore, there are  only 6 physical
degrees of freedom. Due to the relations
$\{r_i,\chi\}=0$, $\{\Pi_i,\chi\}\approx 0$,
the variables $r_i$ and $\Pi_i$ give such a
set of physical degrees of freedom.
The Hamiltonian of the initial CM
system has a form $H=\frac{1}{2}{\bf P}^2$,
and the nontrivial Poisson brackets for the initial system
are $\{r_i,P_j\}=\delta _{ij}$,
$\{P_i,P_j\}=\nu r^{-3}\epsilon _{ijk}r_k$.
Taking into account that for the system (\ref{lfspin})
the Poisson brackets
of  the gauge-invariant (physical) variables are
$
\{\Pi_i,\Pi_j\}=-\frac{\bf rs}{r^4}\epsilon_{ijk}r_k\approx
\nu r^{-3}\epsilon _{ijk}r_k$
and
$
\{\Pi_i,r_j\}=\delta_{ij},
$
one concludes that the variables
$\Pi _i$ are the analogs of
$P_i=\dot{r}_i$, and,
as a result, the dynamics generated by the Hamiltonian
(\ref{ht}) for the gauge-invariant variables is exactly
the same as the dynamics of the initial CM
system.

Therefore, we conclude that the CM system
can alternatively be interpreted as a free particle
of fixed spin defined by the value of the
CM coupling constant $\nu$
with translational and spin degrees of freedom interacting
via the helicity constraint and the ${\bf Ls}$-coupling term.

\section{CM  and (2+1)D anyon}
The observed effectively free
nature of the CM dynamics
allows ones also to trace out  the origin and details of
the known formal  relation  of the CM system
to the (2+1)-dimensional free relativistic anyon
\cite{sch,any0,skag,tor5}.

In (2+1)-dimensions, spin is a (pseudo)scalar and,
as a consequence, the anyon of fixed spin has the same
number
of degrees of freedom as a spinless free massive particle
\cite{any0,tor5,tor1}.
The relationship between the
CM and anyon can be understood better
within the framework of the canonical description of these
two systems.
As we have seen, the CM system essentially
is a reduced E(3) system. In the case of anyon,
the E(3) group is changed for the Poincar\'e group
ISO(2,1).
The translation generators of the corresponding groups
are $n_i$  and $p_\mu$, the latter being the
energy-momentum vector of the anyon, and the
corresponding Casimir central elements are fixed
by the relations ${\bf n}^2=1$ and
$
p^2+m^2\approx 0,
$
where $m$ is a mass of the anyon.
The rotation (Lorentz) generators are given by ${\bf J}$
and by
\beq
\label{jxp}
{\cal J}_\mu=\epsilon_{\mu\nu\lambda}x^\nu p^\lambda
+J_\mu,
\eeq
where $J_\mu$  are the translation invariant
$so(2,1)\sim sl(2,R)$
generators satisfying  the algebra
\beq
\{J_\mu,J_\nu\}=\epsilon_{\mu\nu\lambda}J^\lambda,
\eeq
and subject to the relation
$J_\mu J^\mu=-\gamma^2=const$.
Representation (\ref{jxp}) for the
anyon total angular momentum vector is,
obviously, the analog of the relation
${\bf J}={\bf L}+{\bf s}$
appearing under interpretation of the CM
system as a particle with spin.
In the CM system, the second Casimir element
of E(3) is fixed either strongly, ${\bf Jn}=-\nu$,
or in the form of the weak relation
${\bf Jn}+\nu\approx 0$ being equivalent to the
helicity constraint under treating the CM
as the particle with spin.
In the anyon model, spin also can be fixed either
strongly, ${\cal J}p=Jp=-\gamma m$, or in the form of the
weak (constraint) relation
$
\chi_a\equiv Jp+\gamma m\approx 0
$
\cite{tor5}.
When the helicity is fixed strongly,
for the CM system the
symplectic form
has a nontrivial contribution describing
noncommuting quantities $P_i$ being
the components of the charge's velocity:
$\omega=dP_i\wedge dr_i+\frac{\nu}{2r^3}\epsilon_{ijk}
r_idr_j\wedge
dr_k$.
In the anyon case, the strong spin fixing
gives rise to the nontrivial Poisson structure
for the particle's coordinate's \cite{tor5},
\beq
\label{anyx}
\{x_\mu,x_\nu\}=
\alpha (-p^2)^{-3/2} \epsilon_{\mu\nu\lambda}p^
\lambda.
\eeq
On the other hand, when we treat the CM
system as a particle with spin (helicity is fixed weakly),
one can work in terms of the canonical symplectic
structure for the charge's coordinates and momenta,
$\omega=dp_i\wedge dr_i+\omega_s$,
but the canonical momenta
${\bf p}$ are not physical observables due to their
non-commutativity with the helicity constraint,
whereas the gauge-invariant extension of $p_i$
given by the variables $\Pi _i$
plays the role of
the non-commuting quantities $P_i$.
Exactly the same picture takes place in
the case of anyon when its spin is fixed
weakly: the coordinates $x_\mu$ commute in this case,
$\{x_\mu,x_\nu\}=0$,
but they have nontrivial Poisson brackets
with the spin constraint
$
\chi_a
$,
whereas their gauge-invariant extension,
$X_\mu=x_\mu+\frac{1}{p^2}\epsilon_{\mu\nu\lambda}p
^\nu J^\lambda$,
$\{X_\mu,\chi_a\}\approx 0$,
are non-commuting and
reproduce the Poisson bracket relation (\ref{anyx})
\cite{tor5}.
Like in the anyon case, the advantage
of the extended formulation for the
CM system (when we treat it as a particle
with spin)  is in the existence of canonical
charge's coordinates $r_i$ and momenta $p_i$.
Within the initial minimal formulation
(given in terms of $r_i$ and gauge-invariant
variables $P_i$), the canonical momenta $p_i$
are reconstructed from $P_i$ only locally,
$p_i=P_i+eA_i$, due to the global
Dirac string singularities hidden in
the monopole vector potential.
Having in mind the gauge invariant nature
and non-commutativity
of $P_i$ or their analogs $\Pi_i$ from the extended
formulation, we conclude that they, like
anyon coordinates $X_\mu$,
are the charge-monopole's analogs
of the Foldy-Wouthuysen  coordinates of the
Dirac particle \cite{tor5,zit}.

\section{Fermion-monopole supersymmetry}

Now we apply the discussed  effectively free
dynamics of the CM  for clarifying the
nature of the non-standard supersymmetry of the
fermion-monopole.
As it was  observed for the first time
in ref. \cite{hid},
in addition to the standard
$N=1/2$ supersymmetry \cite{hv},
the fermion-monopole system has
the hidden supersymmetry of the non-standard form
characterized by the supercharge
anticommuting for the nonlinear operator
equal to the
square of the total angular momentum operator
shifted for the constant.
The nonlinear supersymmetry of the similar form
 was also revealed in the 3D $P,T$-invariant systems
of relativistic fermions \cite{gps} and Chern-Simons fields
\cite{np}
as well as  in the pure parabosonic systems and Calogero-
like
systems with exchange interaction \cite{nparab}.
The relation of the nonlinear supersymmetry
to the quasi-exactly solvable systems
via the quantum anomaly
was investigated recently in ref. \cite{nsusy}.

To clarify the origin of such a supersymmetry
and its structure, let us consider following ref. \cite{mp2}
a 3D free spin-1/2 nonrelativistic
particle given by the
classical Lagrangian
\beq
\label{lan0}
L=\frac{1}{2} \dot{\bsymb{r}}{}^2+\frac{i}{2}\bsymb{
\psi}\dot{\bsymb{\psi}}.
\eeq
The corresponding Hamiltonian is
$H=\frac{1}{2}\bsymb{p}^2$,
and nontrivial Poisson-Dirac brackets
for Grassmann variables
are
$
\{\psi_i,\psi_j\}=-i\delta_{ij}.
$
The set of even integrals of motion
is given by the vectors
$\bsymb{p}$,
$\bsymb{L}=
\bsymb{r}\times \bsymb{p}$ and $\bsymb{K}=
\bsymb{p}\times \bsymb{L}$,
which have a nonzero projection on a unit of Grassmann
algebra, and by the nilpotent spin vector
$\bsymb{S}=
-\frac{i}{2}\bsymb{\psi}\times\bsymb{\psi}$
generating rotations of odd Grassmann variables $\psi_i$.
The vector $\bsymb{K}$ is the analog of the Laplace-
Runge-Lentz vector of the Kepler system, which together
with
$\bsymb{p}$ and $\bsymb{L}$ constitute a
non-normalized basis of orthogonal vectors
forming a nonlinear algebra
with respect to the Poisson brackets.
Projecting the odd vector integral $\bsymb{\psi}$
onto the even vector integrals of motion,
we get three odd scalar integrals of motion (supercharges)
\beq
\label{susysc}
Q_1=\bsymb{p\psi},\quad
Q'_2=\bsymb{L\psi},\quad
Q_3=\bsymb{K\psi}.
\eeq
The supercharge $Q_1$ is a ``square root from the
Hamiltonian",
$\{Q_1,Q_1\}=-2iH$.
It has zero bracket
with the supercharge $Q_3$, $\{Q_1,Q_3\}=0$,
but $Q_1$ and $Q_3$ have nontrivial brackets with
$Q'_2$.
One can find the linear combination of
the odd scalar integrals $Q'_2$ and $i\epsilon_{ijk}\psi_i
\psi_j\psi_k$
having zero brackets with other two supercharges,
\beq
\label{q2}
Q_2=\bsymb{L\psi}-\frac{i}{3}(\bsymb{\psi}\times
\bsymb{\psi})\cdot\bsymb{\psi}.
\eeq
Finally, we get the set of three scalar supercharges,
$\{J_i,Q_a\}=0,$
$a=1,2,3$,
forming together with $H$ and $\bsymb{J}$
the nonlinear superalgebra
\beq
\label{clalg}
\{Q_a,Q_b\}=-iA_a\delta _{ab},
\eeq
where
$A_1=2H$,
$A_2=\bsymb{J}^2$,
$A_3=A_1A_2$,
and
$\bsymb{J}=\bsymb{L}+\bsymb{S}$
is the total angular momentum vector.
The scalar supercharges
satisfy also the algebraic relation
\beq
Q_aQ_b=iB_c\epsilon_{abc}\tilde{Q}_c,
\label{sualg}
\eeq
where $B_1=A_2$, $B_2=A_1$, $B_3=1$
and
$\tilde{Q}_a$ means $Q_a$ with
odd vector $\bsymb{\psi}$ changed for even $\bsymb{S}
$.
Since $\bsymb{J}^2\cdot\tilde{Q}_1=\bsymb{L}^2\cdot
\tilde{Q}_1$,
relations (\ref{sualg}) reflect
the analogous relations between the even
vector integrals $\bsymb{p}$, $\bsymb{L}$ and
$\bsymb{K}$:
$
\bsymb{p}\times \bsymb{L}=\bsymb{K},
$
$
\bsymb{p}\times \bsymb{K}=-2H\cdot\bsymb{L},
$
$
\bsymb{L}\times
\bsymb{K}=\bsymb{L}^2\cdot \bsymb{p}.
$
Taking into account the equalities
$\bsymb{J}^2=2H\cdot\Delta$
and
$
\{\Delta,H\}=\{\Delta,Q_a\}=0,
$
where
$\Delta=\bsymb{r}^2_\perp+
(\bsymb{LS})\cdot{H}^{-1}$,
$\bsymb{r}_\perp=\bsymb{r}-\bsymb{p}(\bsymb{pr})
\cdot\bsymb{p}^{-2}=
\bsymb{K}\cdot\bsymb{p}^{-2}$,
one can transform
(at $\bsymb{p}^2\neq 0$, $\bsymb{K}^2\neq 0$)
the set of scalar supercharges
$Q_a$ into the set
\beq
\bar{Q}_a=C_a Q_a
\label{bq}
\eeq
with $C_1=1$,
$C_2=\Delta^{-1/2},$
$C_3=(2H\Delta)^{-1/2}$.
This set of (nonlinearly) transformed scalar
supercharges gives rise to the $N=3/2$ linear superalgebra
of the standard
form,
\beq
\label{susys}
\{\bar{Q}_a,\bar{Q}_b\}=-2iH\delta_{ab}.
\eeq

The quantum analogs of the odd variables are realized
via the Pauli matrices,
$\hat{\psi}_i=\sigma_i/{\sqrt{2}}$,
and the quantum spin vector $\hat{\bf S}$ is
proportional (``parallel") to $\hat{\bsymb{\psi}}$:
$\hat{\bf S}=\bsymb{\sigma}/2=
\hat{\bsymb{\psi}}/\sqrt{2}$.
Classically this property is reflected
in the relation $\bsymb{\psi}\times {\bf S}=0$.
To construct the quantum analogs of the supercharges
$Q_a$ in the form of Hermitian operators,
we choose the natural prescription ${\bf p}\times {\bf L}
\rightarrow
\frac{1}{2}(\hat{\bf p}\times\hat{\bf L}+(\hat{\bf p}\times
\hat{\bf L})^\dagger)$,
and get
\beq
\label{quq00}
\hat{Q}_1=\hat{\bf p}\hat{\bsymb{\psi}},\quad
\hat{Q}_2=\hat{\bf L}\hat{\bsymb{\psi}}+
\frac{1}{\sqrt{2}},
\eeq
\beq
\label{quq0}
\hat{Q}_3=(\hat{\bf p}
\times\hat{\bf L})\hat{\bsymb{\psi}}-i\hat{Q}_1.
\eeq
These operators are Hermitian and
satisfy the quantum relations
\beq
\hat{Q}_a\hat{Q}_b=\frac{1}{2}\hat{A}_a\delta_{ab}
+
\frac{i}{\sqrt{2}}\hat{B}_c\epsilon_{abc}\hat{Q}_c,
\label{qususy00}
\eeq
where
$\hat{A}_1=2\hat{H},$
$\hat{A}_2=\hat{\bf J}{}^2+\frac{1}{4},$
$\hat{A}_3=\hat{A}_1\hat{A}_2,$
$\hat{B}_1=\hat{A}_2,$
$\hat{B}_2=\hat{A}_1,$
$\hat{B}_3=1.$
The symmetric part of relations (\ref{qususy00})
is the nonlinear $N=3/2$ superalgebra  being the exact
quantum analog
of classical relations (\ref{clalg}),
whereas,
with taking into account the above mentioned
relation between $\hat{S}_i$ and $\hat{\psi}_i$, we find
that
the antisymmetric part of (\ref{qususy00}) corresponds to
the
classical relations (\ref{sualg}).

Due to the relation $\hat{Q}_3=-i\sqrt{2}\hat{Q}_1\hat{
Q}_2$,
it seems that one could interpret $\hat{Q}_1$ and
$\hat{Q}_2$
as the ``primary" supercharge operators and $\hat{Q}_3$
as the
``secondary" operator. But such an interpretation
is not correct. Indeed,
on the one hand, the relations (\ref{qususy00})
can be treated as a quantum generalization of the
symmetric in indexes relations for the three Clifford algebra
generators $\sigma_i=\sqrt{2}\hat{\psi}_i$,
$\sigma_i\sigma_j=\delta_{ij}+i\epsilon_{ijk}\sigma_k$,
and we remember that the operators $\hat{\psi}_i$
are the quantum analogs of the set of
the classical odd integrals of motion $\psi_i$
forming a 3D vector.
On the other hand, the antisymmetric part of
(\ref{qususy00})
is a reflection of the
cyclic classical relations
(\ref{sualg}).
It is worth noting also that
we can pass from the set of odd integrals
(\ref{bq}) (constructed at
$\bsymb{p}^2\neq 0$, $\bsymb{K}^2\neq0$)
to the integrals
\beq
\label{trgr}
\xi_a=\bar{Q}_a\cdot(2H)^{-1}.
\eeq
Then, due to the relations $\{\xi_a,\xi_b\}=-i\delta_{ab}$,
one can treat the transition from the integrals $\psi_i$
to the integrals $\xi_a$ as a simple
canonical transformation for the odd
sector of the phase space (note, however, that $\xi_a$
have nontrivial brackets with even variables).
Formally, the same unitary in the odd sector transformations
can be realized at the quantum level.
So, it is natural to treat
all the three supercharge operators $\hat{Q}_a$
on the equal footing.

Let us consider the fermion particle
of charge $e$ in
arbitrary time-independent magnetic field
$\bsymb{B}(\bsymb{r})=
\bsymb{\nabla}\times
\bsymb{A}(\bsymb{r})$. At the Hamiltonian level
this corresponds to the change of the canonical
momentum vector $\bsymb{p}$ for the vector
$\bsymb{P}=\bsymb{p}-e\bsymb{A}$,
$\{P_i,P_j\}=e\epsilon_{ijk}B_k$.
By projecting the odd vector variable $\bsymb{\psi}$
onto $\bsymb{P}$, we get the scalar
$Q_1=\bsymb{P\psi}$.
Identifying the bracket $\frac{i}{2}\{Q_1,Q_1\}$
as the Hamiltonian,
\beq
\label{hmag}
H=\frac{1}{2}\bsymb{P}^2-e\bsymb{BS},
\eeq
$Q_1$ is automatically the odd integral of motion
(supercharge).
However, now, unlike the case of a free particle,
either even vectors $\bsymb{P}$, $\bsymb{L}=\bsymb{r}
\times\bsymb{P}$,
$\bsymb{P}\times\bsymb{L}$, $\bsymb{S}$
or odd vector $\bsymb{\psi}$ are not integrals of motion,
whereas the odd scalar $(\bsymb{\psi}\times\bsymb{\psi})
\cdot\bsymb{\psi}$
is conserved. Having in mind the analogy with
the free particle case, let us check
other odd scalars for their possible conservation.
First, it is worth noting that the quantum
relation $\hat{\bsymb{S}}=\frac{1}{\sqrt{2}}
\hat{\bsymb{\psi}}$ is reflected classically
also in the  identical evolution,
$\dot{\bsymb{\psi}}=e\bsymb{\psi}\times
\bsymb{B}$, $\dot{\bsymb{S}}=e\bsymb{S}\times
\bsymb{B}$.
Like the projection of the odd vector
$\bsymb{\psi}$,
the projection of $\bsymb{S}$  on $\bsymb{P}$
is conserved,
i.e. as in a free case, $\tilde{Q}_1=\bsymb{PS}$ is the
integral
of motion, but generally
$\frac{d}{dt}(\bsymb{SL})=e(\bsymb{S}\times\bsymb{P
})\cdot
(\bsymb{r}\times\bsymb{B})\neq 0$.
The scalar $\tilde{Q}{}'_2=\bsymb{SL}$
is the integral of motion only if
$\bsymb{B}=f(r)\cdot\bsymb{r}$, $r=\sqrt{\bsymb{r}^2}
$.
This corresponds exactly to the case of the monopole field,
for which $f(r)=gr^{-3}$ ($\bsymb{r}\neq 0$, $g=const$)
is fixed by the condition
$\bsymb{\nabla B}=0$, and from now on, we restrict the
analysis
by the fermion-monopole system.
Though in this case
the brackets
$\{L_i,L_j\}=\epsilon_{ijk}(L_k+\nu n_k)$
are different from the corresponding brackets for the free
particle, nevertheless the odd scalar
$Q'_2=\bsymb{L\psi}$
is the integral of motion.
One can check that
the direct analog of the free particle's supercharge
(\ref{q2}) has zero bracket with $Q_1$
and that
\beq
\label{q2mon}
\{Q_2,Q_2\}=-i(\bsymb{\cal J}^2-\nu^2).
\eeq
Here $\bsymb{\cal J}=
\bsymb{J}+\bsymb{S}$,
$\bsymb{J}=\bsymb{L}-\nu {\bf n}$,
is the conserved angular momentum vector
of the fermion-monopole system,
whose components form $su(2)$ algebra,
$\{{\cal J}_i,{\cal J}_j\}=\epsilon_{ijk}{\cal J}_k$,
and generate rotations.
The scalar $Q_3=(\bsymb{P}\times \bsymb{L})
\bsymb{\psi}$
is also the integral of motion and
in the fermion-monopole case
the classical relations of the form
(\ref{clalg}), (\ref{sualg})
take place with the change
$\bsymb{J}^2\rightarrow
\bsymb{\cal J}^2-\nu^2$ and with the Hamiltonian
given by Eq. (\ref{hmag}). Like for the free particle,
the superalgebra can be reduced to the standard linear
form (\ref{susys}) via the nonlinear transformation
(\ref{bq}),
for which in the present case we proceed from the relation
$\bsymb{\cal J}^2-\nu^2=2H\cdot\Delta$
with $\Delta=\tilde{\bsymb{r}}{}^2+
\bsymb{LS}\cdot H^{-1}$,
$\tilde{\bsymb{r}}{}^2=\bsymb{r}^2-
(\bsymb{Pr})^2\cdot(2H)^{-1}$.
The quantity  $\tilde{\bsymb{r}}{}^2$
is the integral of motion which in the case of
the scalar charged particle ($\psi_i=0$)
in the field of monopole
gives a minimal charge-monopole distance
in the point of perihelion:
$r_{min}=\sqrt{\tilde{\bsymb{r}}{}^2}$ \cite{mp1}.

Constructing the quantum analogs of the supercharges
in the same way as in the free particle case,
we get the supercharge operators of the
form (\ref{quq0}) with
$\hat{\bsymb{p}}$ changed for $\hat{\bsymb{P}}$.
They satisfy the set of (anti-)commutation relations
of the same form
(\ref{qususy00})
with
$\hat{A}_2=\hat{\bsymb{ J}}{}^2+\frac{1}{4}$
changed
for $\hat{\bsymb{\cal J}}{}^2-\nu^2+\frac{1}{4}$,
where $\nu$, as in the case of the CM system,
is subject to the Dirac quantization
condition, $2\nu\in \ZZ$.

The present analysis shows that the set of
supercharge operators $\hat{Q}_1$ and $\hat{Q}_2$
found in ref. \cite{hid}
has to be extended by the scalar integral $\hat{Q}_3$,
and these three odd operators together with even
operators $\hat{H}$ and $\hat{\bsymb{\cal J}}$
form the described nonlinear $N=3/2$ superalgebra.
As we have seen, this
nonlinear supersymmetry of the fermion-monopole system
has the nature of the  free fermion particle's supersymmetry
generated by the supercharges represented in a scalar form.

{}Comparing the fermion-monopole
Hamiltonian (\ref{hmag}) 
(with $\bsymb{B}=\nu \bsymb
{r}\cdot r^{-3}$)
with the free fermion particle Hamiltonian
$H=\frac{1}{2}\bsymb{p}^2$, it seems that they have
rather different structure, but this is not so,
and their similarity can be revealed, like in the case
of the CM system,
by separating the even phase space coordinates into the
radial and angular ones.
In terms of such variables, the Hamiltonian of the
fermion-monopole system is
\beq
H=\frac{1}{2}P_r^2+\frac{\bsymb{\cal J}^2-\nu^2}{2r^
2}
-\frac{\bsymb{LS}}{r^2},
\eeq
with $\bsymb{\cal J}=\bsymb{J}+\bsymb{S}$.
The case  of the free fermion corresponds to $\nu=0$,
and its Hamiltonian takes the similar form
$
H=\frac{1}{2}p_r^2+\frac{1}{2r^2}\bsymb{J}^2
-\frac{1}{r^2}\bsymb{LS},
$
where
$\bsymb{J}=\bsymb{L}+\bsymb{S}$.

It is interesting to
look at the fermion-monopole supersymmetry
from the point of view
of the reduction of the system to the
spherical geometry\footnote{Some aspects
of the CM system in a spherical
geometry were discussed in refs. \cite{khar,mp1}.}.
To this end we first note that the bracket of the supercharge
$Q_2$
with itself can be represented in the form
\beq
\label{q2red}
i\{Q_2,Q_2\}=2r^2\left(H-
\frac{1}{2}P_r^2+\frac{i}{r}Q_1(\bsymb{\psi n})\right),
\eeq
and the supercharge $Q_3$ can be reduced to the
equivalent form
\beq
\label{q3red}
Q_3=2Hr\left(\bsymb{\psi n}-Q_1\frac{P_r}{2H}\right).
\eeq
The reduction of the fermion-monopole system
to the spherical geometry
can be realized
by introducing into the system the classical relations
\beq
\bsymb{r}^2-\rho^2=0,\quad
P_r=0,\quad
\bsymb{\psi n}=0,\label{sfera1}
\eeq
which have to be treated as the set of second class
constraints
with  $\rho\neq0$ being a constant, and for simplicity we fix
it in the form $\rho=1$.
The relations (\ref{q2red}) and (\ref{q3red})
allow us to observe directly that
the described $N=3/2$ nonlinear fermion-monopole
supersymmetry is transformed into  the $N=1$
supersymmetry
of the standard linear form in the case
of reduction (\ref{sfera1}).
Indeed, after reducing the fermion-monopole system
onto the surface of even second class constraints
(\ref{sfera1}),
we find that the structure of the supercharge $Q_3$ is
trivialized and takes the form of
the odd scalar $\bsymb{\psi n}$ multiplied by $2H$.
Two other supercharges $Q_1$ and $Q_2$ after
such a reduction take the form of  linear combinations
of the odd vector $\bsymb{\psi}$ projected
on the vectors $\bsymb{J}+\nu\bsymb{n}$ and
$\bsymb{ J}\times\bsymb{n}$ orthogonal to
$\bsymb{n}$.
Then taking into account the odd second class constraint
(\ref{sfera1}) results in eliminating the supercharge $Q_3$
and in reducing the bracket
(and corresponding anticommutator at the quantum level)
of the supercharge $Q_2$ to $2h$,
where $h$ is the reduced Hamiltonian,
\beq
\label{hreds}
h=\frac{1}{2}(\bsymb{\cal J}^2-\nu^2).
\eeq
In other words, the supersymmetry of the fermion-
monopole
system in spherical geometry is
reduced to the standard linear $N=1$ supersymmetry
characterized by two supercharges anticommuting
for the Hamiltonian.
More explicitly, after reduction to the surface of the
second class constraints, the radial variables
$r$ and $P_r$ are eliminated from the theory.
The even variables can be  represented by
the total angular momentum $\bsymb{\cal J}$ and by the
unit
vector
$\bsymb{n}$ having the nontrivial Dirac brackets
coinciding with corresponding initial Poisson brackets,
$\{{\cal J}_i,{\cal J}_j\}^*=\epsilon_{ijk}{\cal J}_k$,
$\{{\cal J}_i,n_j\}^*=\epsilon_{ijk}n_k$.
The odd variables $\psi_i$ satisfy
the relation
$\bsymb{\psi n}=0$
which has to be treated as a
strong
equality, and their nontrivial Dirac brackets
are $\{\psi_i,\psi_j\}^*=-i(\delta_{ij}-n_in_j)$,
$\{{\cal J}_i,\psi_j\}^*=\epsilon_{ijk}\psi_k$.
The even and odd variables are subject
also to the relation
$\bsymb{{\cal J}n}=-\nu+iq_1q_2\cdot(2h)^{-1}$,
where $h$ is given by Eq. (\ref{hreds})
and
\beq
q_1=(\bsymb{\cal J}\times\bsymb{n})\bsymb{\psi},
\quad
q_2=\bsymb{{\cal J}\psi}
\eeq
are the supercharges $Q_1$ and $Q_2$
reduced to the surface (\ref{sfera1}).
With the listed Dirac brackets, one can easily
check that now the reduced supercharges
$q_\mu$, $\mu=1,2$,
satisfy
the superalgebra of the standard $N=1$ supersymmetry:
$\{q_\mu,q_\nu\}^*=-2i\delta_{\mu\nu}h$,
$\{q_\mu,h\}=0$.

\vskip0.3cm
{\bf Acknowledgements}
\vskip3mm
The work was supported in part by the
grants 1980619  and 1010073
from FONDECYT (Chile)
and by DICYT (USACH).
It is a pleasure to acknowledge the stimulating friendly
atmosphere of the D. V. Volkov Memorial Conference
and the very kind hospitality by the local organizers.

\end{document}